\documentclass[]{spie}  

\usepackage{amsmath,amsfonts,amssymb}
\usepackage{graphicx}
\usepackage[colorlinks=true, allcolors=blue]{hyperref}
\usepackage{aas_macros}

\usepackage{subfigure} 
\usepackage{multicol}

\title{Packing the sky: coverage optimization and evaluation for large telescope arrays}

\author{Nathan W. Galliher}
\author{Nicholas M. Law}
\author{Hank Corbett}
\author{Ramses Gonzalez}
\author{Lawrence Machia}
\author{Alan Vasquez Soto}
\affil{Department of Physics and Astronomy, University of North Carolina at Chapel Hill, Chapel Hill, NC 27599-3255, USA}

\authorinfo{Further author information: (Send correspondence to N.W.G.)\\N.W.G.: E-mail: nathan.galliher@unc.edu}

\begin{document} 
\maketitle

\begin{abstract}
Recent advancements in low-cost astronomical equipment, including high-quality medium-aperture telescopes and low-noise CMOS detectors, have made the deployment of large optical telescope arrays both financially feasible and scientifically interesting. The Argus Optical Array is one such system, composed of 900 eight-inch telescopes, which is planned to cover the entire night sky in each exposure and is capable of being the deepest and fastest Northern Hemisphere sky survey. With this new class of telescope comes new challenges: determining optimal individual telescope pointings to achieve required sky coverage and overlaps for large numbers of telescopes, and realizing those pointings using either individual mounts, larger mounting structures containing telescope subarrays, or the full array on a single mount. In this paper, we describe a method for creating a pointing pattern, and an algorithm for rapidly evaluating sky coverage and overlaps given that pattern, and apply it to the Argus Array. Using this pattern, telescopes are placed into a hemispherical arrangement, which can be mounted as a single monolithic array or split into several smaller subarrays. These methods can be applied to other large arrays where sky packing is challenging and evenly spaced array subdivisions are necessary for mounting.
\end{abstract}

\keywords{Argus Optical Array, telescope, large arrays, wide-field surveys, sky coverage}

\section{INTRODUCTION}
\label{sec:intro}  
The Evryscopes are a northern and southern pair of all-sky telescope arrays comprised of 24 6.1-cm aperture telescopes on a shared mount. By sacrificing survey depth and sky sampling for extreme field of view (FoV) and high-cadence observations, each Evryscope monitors over 8,000 square degrees in every two-minute exposure\cite{law_2015, ratzloff_2019}. The Evryscopes observe the sky in a series of ``ratchets,'' tracking the sky for two hours before ratcheting back to start another set of observations. These telescopes have been used in conventional surveys searching for long-term variability (e.g.\ Ref.~\citenum{ratzloff2019variables, ratzloff_2020_hot, galliher_2020}), while their innovative design allows for a new parameter space to be explored: the short-timescale transient sky. The Evryscopes have detected rapid transient events ranging from flares and superflares (e.g.\ Ref.~\citenum{howard_superflare, howard_evryflare_1, howard_evryflare_2, howard_evryflare_3, glazier_trappist}) to optical reflections of space debris (e.g.\ Ref.~\citenum{corbett_2020}).

The Argus Optical Array is a proposed next-generation system designed for deep, high-cadence, and all-sky observations\cite{law_2022}. The Argus Array will be comprised of 900 mass-produced medium-aperture telescopes and high-speed sensors. Argus, just like the Evryscopes, takes advantage of recent technological advancements that make it more cost effective to build an array of many smaller telescopes, compared to a single large-aperture telescope\cite{LAST}. The array will have the collecting area of a 5-m telescope, an étendue approaching VRO (LSST), and the ability to follow the evolution of every $m_g<23.6$ time-variable source across the sky simultaneously. 

\begin{figure}
    \centering
    \includegraphics[width=0.98\textwidth]{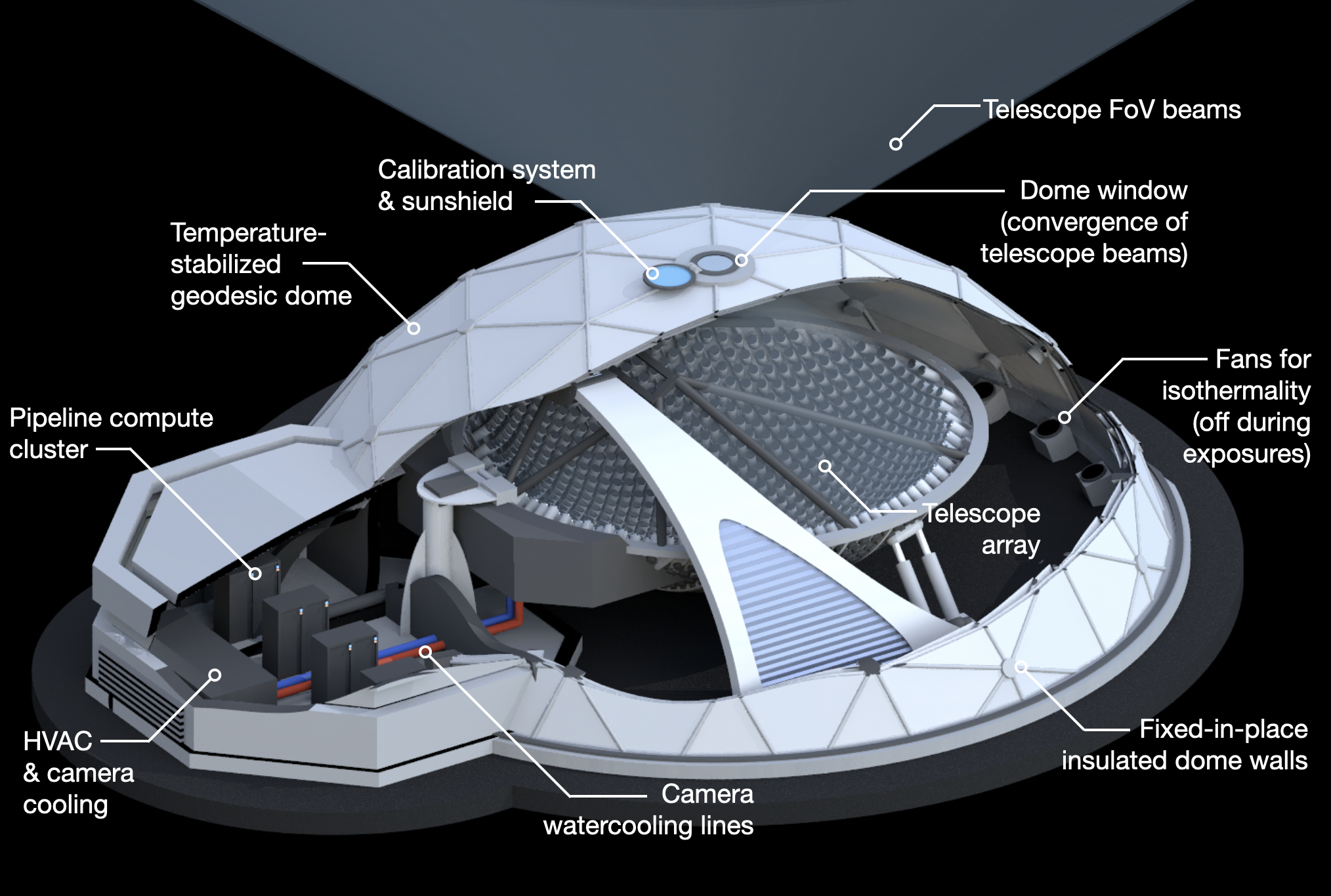}
    \caption{The Argus Array concept. Argus will contain 900 telescopes mounted in a shared hemispherical ``bowl.'' The telescopes point radially inward toward the center of the bowl through a window. The array tracks the sky during 15 minute ratchets, before ratcheting back to start a new set of observations. Argus is placed in a thermally controlled, weather-sealed building.}
    \label{fig:argus}
\end{figure}

The Argus Array is currently in the prototyping phase. The soon-to-be deployed prototype, called the Argus Pathfinder, will be an array of 38 telescopes that monitor a stripe of declination (dec.), from -20$^{\circ}$ to 72$^{\circ}$, for 15 minutes at a time; over the course of an observing night Pathfinder will build up observations for most of the visible sky\cite{argus_spie}. The Argus Array and Pathfinder telescopes will be rigidly mounted in hemispherical ``bowl'' mounts, that will track the sky in 15-minute ratchets. The bowls will be weather-sealed inside of a building, with the virtual center of the hemispherical bowl positioned on a window through which all of the telescopes look (Fig.~\ref{fig:argus}).

The Evryscopes, Argus, and Pathfinder all share a common design constraint that greatly reduces their mechanical complexity: the telescopes in each of the arrays are rigidly coupled to their shared mounts. The mounts hold the arrays in a hemispherical arrangement, so that the all telescope apertures lie on the surface of a sphere, mimicking the shape of the sky. The telescopes' pointings are normal to this pseudo-surface. Telescopes mounted in the Argus and Pathfinder arrays point radially inward while telescopes mounted in the Evryscope domes point radially outward. Therefore, the location of a telescope in the mount determines its on-sky pointing. By using fixed array pointings, the number of moving parts is reduced by several orders of magnitude. While simplifying the construction and maintenance of the system, this constraint requires careful planning of the ``telescope packing,'' i.e.\ the positioning of telescopes within this hemispherical arrangement. Determining the best packing pattern can be difficult, particularly in next-generation systems, because as the number and size of telescopes increases, and their FoVs decrease, much larger mounting structures are required to achieve continuous sky coverage. For very large arrays, the size of Argus or bigger, where the diameters of the mounts become many tens of feet, it can be beneficial to spread the array over several smaller mounts.  

In this paper, we develop an algorithm for rapidly evaluating the  on-sky performance metric for a large array (Section \ref{sec:eval}) and discuss how it can be used to calculate performance metrics (Section \ref{sec:metrics}). Next, we discuss the on-sky packing geometry for a large array (Section \ref{sec:geometry}), optimizing the array shape (Section \ref{sec:shape}), application of these tools to the Argus Array (Section \ref{sec:app}), and dividing the array into several similar subarrays (Section \ref{sec:multi}). Finally, we summarize conclusions from our work (Section \ref{sec:conc}).

\section{SKY COVERAGE EVALUATION}
\label{sec:cov}
Performance metrics include: total FoV, overlap between telescopes, and limiting magnitudes as the array builds up observations over time. The term ``sky coverage'' refers to the portions of the sky that are observed by the array. 

\subsection{Evaluation Algorithm}
\label{sec:eval}
\begin{figure}
     \centering
    \includegraphics[width=0.98\textwidth]{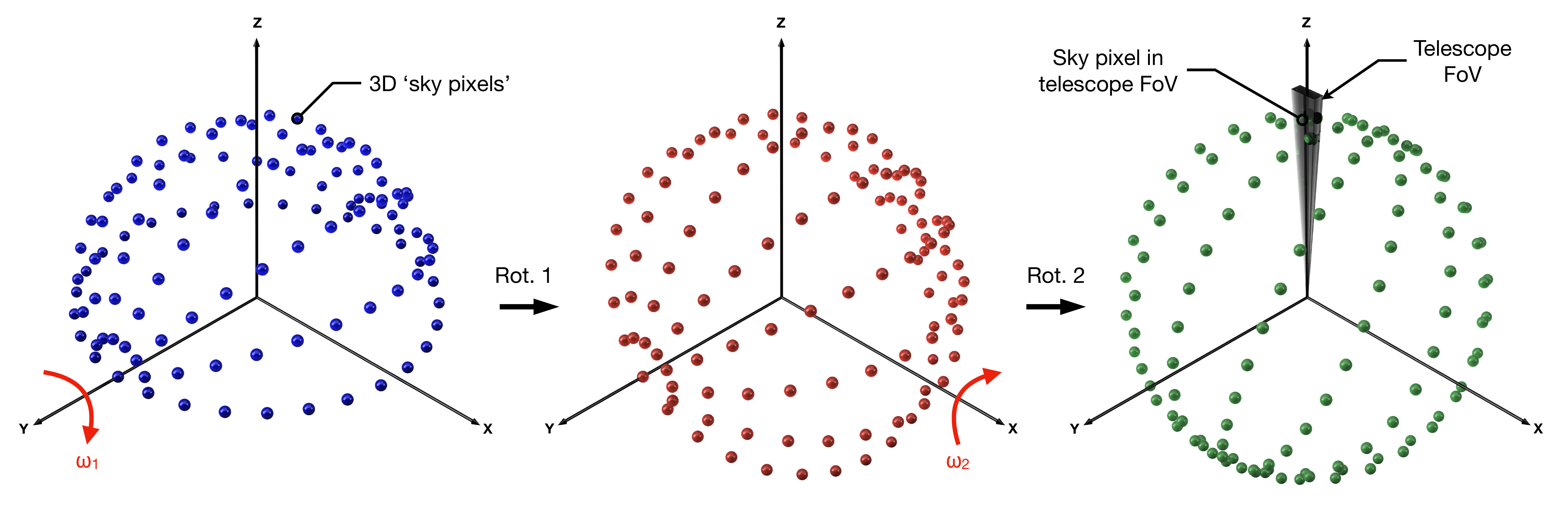}
    \caption{A visual representation of the sky coverage evaluation algorithm. We start by placing a telescope that points along the z-axis. We then create a 2D ``sky pixel'' grid, and convert that to 3D coordinates (blue points, at left). The grid shown here is sparse and only covers half of the sky, for clarity. Next, we perform two rotations, one about the y-axis (red points, center) and then about the x-axis (green points, at right). During this step, it is important to remember to include the minus signs when rotating by $\omega_1 = -$R.A$_{\text{Tele}}$ and $\omega_2 = -$dec$_{\text{Tele}}$. The example shown here is for a telescope pointing in a negative R.A. and positive dec.\ direction. Finally, we check which of the 3D sky pixels intersect the FoV of the telescope, and record the corresponding original 2D sky pixel locations; these pixels in the coverage grid will be observed by the telescope.}
    \label{fig:eval_fig}
\end{figure}

First, we evaluate the sky coverage of an individual telescope at an arbitrary pointing on the sky. The large FoVs of telescopes typically used in arrays make a careful evaluation challenging because the resultant spherical geometry distortion cannot be neglected. Our evaluation algorithm uses the on-sky pointings for all of the telescopes in the array and their FoVs to determine key survey metrics. The telescopes will be oriented so that their rectangular FoVs align with the celestial grid, with either their short or long axis along the direction of R.A., a helpful constraint for the data production pipeline. We define the z-axis to point along the center of the telescope's FoV. The algorithm proceeds as follows:
\begin{enumerate}
    \item We create a mesh grid representing the area of the sky we expect our array to observe. For example, a representative grid for the entire sky would be a mesh grid of values ranging from 0 to 360 in the x-direction (R.A.) and -90 to 90 in the y-direction (dec.). The density of points (``sky pixels'') in this grid will impact the run time; we typically use 1000×1000 points. 
    \item We now convert our 2D sky pixel array into 3D coordinates. We define the coordinates as:
    \begin{equation}
          \begin{gathered}
            X = \cos{(\alpha)}\sin{(\delta + \frac{\pi}{2})}\\
            Y = \cos{(\delta + \frac{\pi}{2})}\\
            Z = \sin{(\alpha)}\sin{(\delta + \frac{\pi}{2})}
          \end{gathered}
    \end{equation}
    with $\alpha$  and $\delta$ being the sky pixel R.A. and dec.\ arrays in radians, respectively. These coordinates place the north celestial pole (NCP) at $(0,-1,0)$.
    \item Next, we rotate our 3D coordinates about two axes; first around the y-axis and then around the x-axis:
    \begin{multicols}{2}
    \noindent
    \begin{equation}
          \begin{gathered}
            X' = X\sin{\omega_1}+Z\cos{\omega_1}\\
            Y' = Y\\
            Z' = X\cos{\omega_1}-Z\sin{\omega_1}
          \end{gathered}
    \end{equation}
    \begin{equation}
          \begin{gathered}
            X'' = X'\\
            Y'' = Y'\cos{\omega_2}-Z'\sin{\omega_2}\\
            Z'' = Y'\sin{\omega_2}+Z'\cos{\omega_2}
          \end{gathered}
    \end{equation}
    \end{multicols}
    where $\omega_1 = -$R.A$_{\text{Tele}}$ and $\omega_2 = -$dec$_{\text{Tele}}$, the negative of the telescope's R.A. and dec.\ pointings in radians, respectively. 
    \item After these rotations, the sky pixel corresponding to the center of the telescope's view will be located at $(0,0,1)$. The positive edge of the telescope's FoV in the R.A. direction will be given by: $\tan{(\frac{\text{FoV}_{\text{x}}}{2}) = \frac{X''}{Z''}}$, where $\text{FoV}_{\text{x}}$ is the known telescope FoV in the x (R.A.) direction, in radians. A similar equality holds true for the dec.\ coverage (y-direction). With this, we can now determine which sky pixels are being observed by the telescope using the following conditions:
    \begin{equation}
          \begin{gathered}
            |\tan({\text{FoV}_{x}}/{2})| > \frac{X''}{Z''}\\
            |\tan({\text{FoV}_{y}}/{2})| > \frac{Y''}{Z''}\\
            Z'' > 0
          \end{gathered}
    \end{equation}
   Wherever these statements hold for the 3D coordinate array, the telescope observes the sky pixels in the 2D mesh grid (Fig.~\ref{fig:eval_fig}). We store these results as a grid of ones and zeroes.
\end{enumerate}

This algorithm is then repeated for every telescope in the array, using batch multiprocessing. By summing all of the resulting grids, we obtain a map of the sky coverage for the total array.

\subsection{Calculating Performance Metrics}
\label{sec:metrics}
For a given telescope arrangement, we use the sky-coverage grid generated above to calculate key performance metrics as follows:

\textbf{Overlap fraction:} The overlap fraction is the percentage of the array's sky coverage that is viewed by more than one telescope in a single exposure. The overlap fraction depends on the value of two weighted sums, one for the number of sky pixels observed by multiple telescopes and one for sky pixels covered by at least one telescope. Both sums use the coverage grid, C. We define the first sum as: $\chi = \sum_{C_{i,j}>1} \cos{\delta_{i,j}}$, and the second sum: $\xi = \sum_{C_{i,j}>0} \cos{\delta_{i,j}}$. The overlap fraction is then: $F={\chi}/{\xi}$.

\textbf{Total FoV coverage:} In terms of the overlap fraction F, the total FoV (in sq.\ deg.\@) is given by: FoV$_{\text{tot}}=\text{FoV}_{x}\times\text{FoV}_{y}\times(1-F)$.

\begin{figure}
     \centering
     \begin{subfigure}
         \centering
         \includegraphics[width=0.48\textwidth]{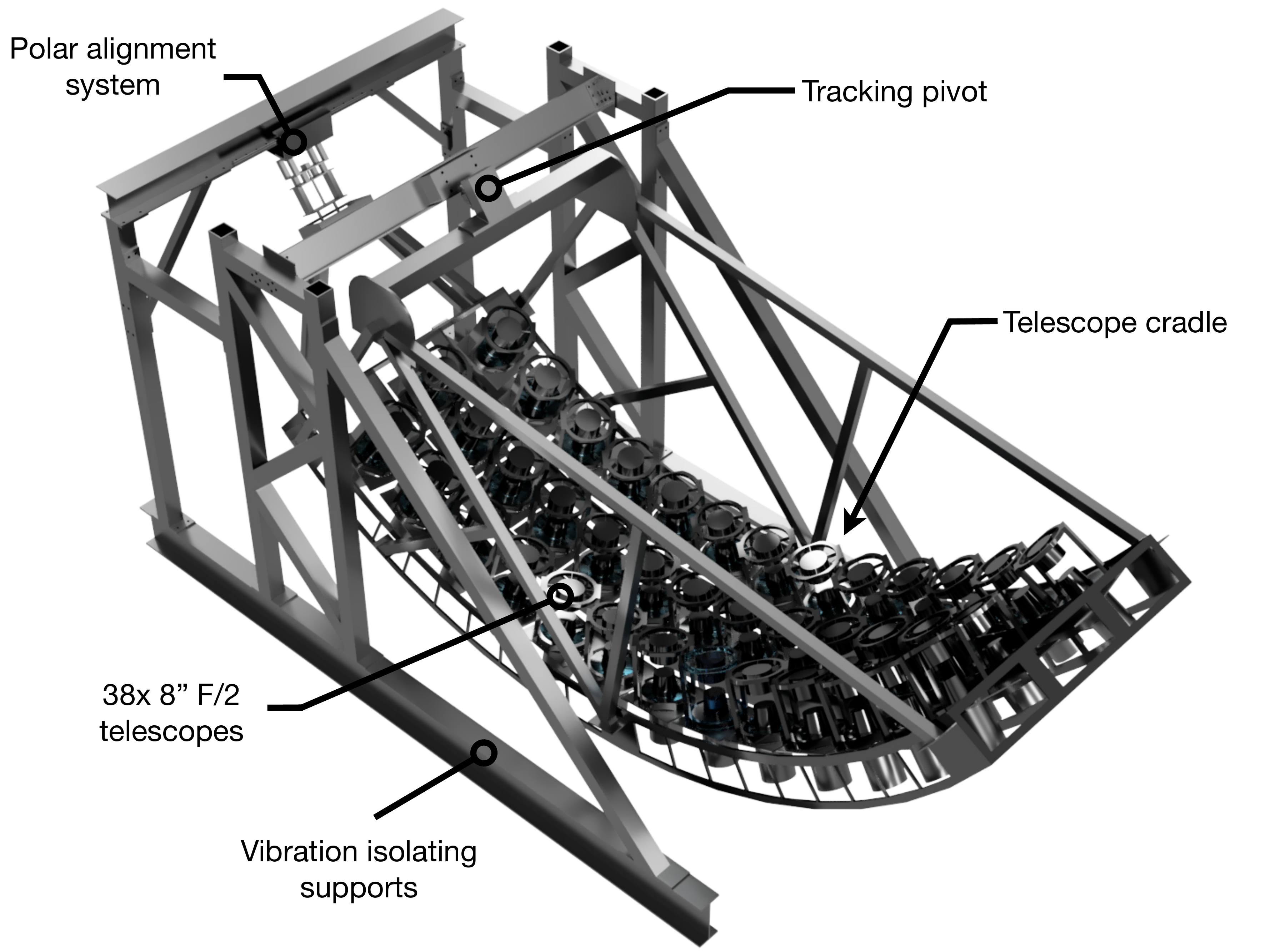}
     \end{subfigure}
     \hfill
     \begin{subfigure}
         \centering
         \includegraphics[width=0.48\textwidth]{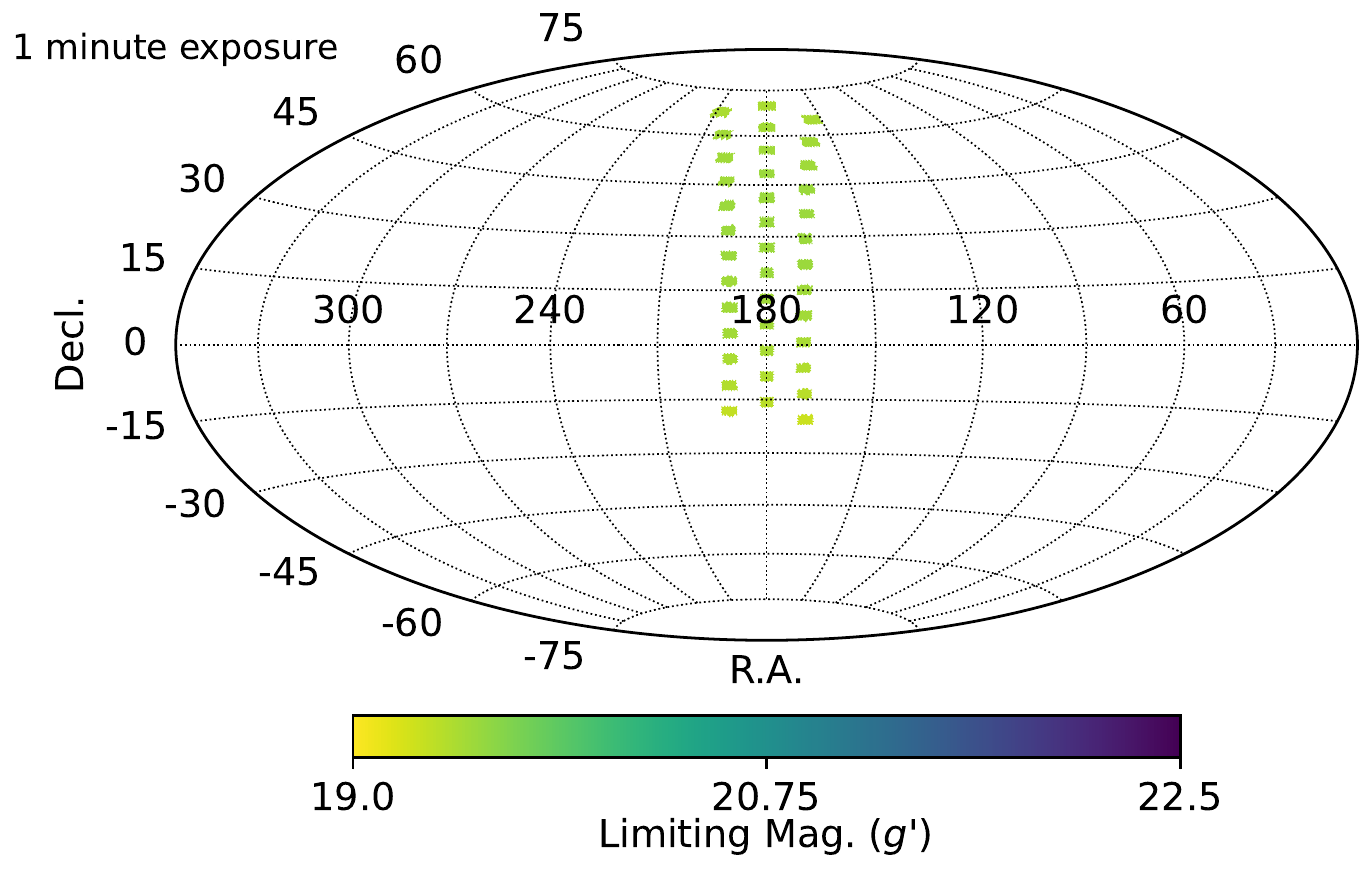}
     \end{subfigure}
     \centering
     \begin{subfigure}
         \centering
         \includegraphics[width=0.48\textwidth]{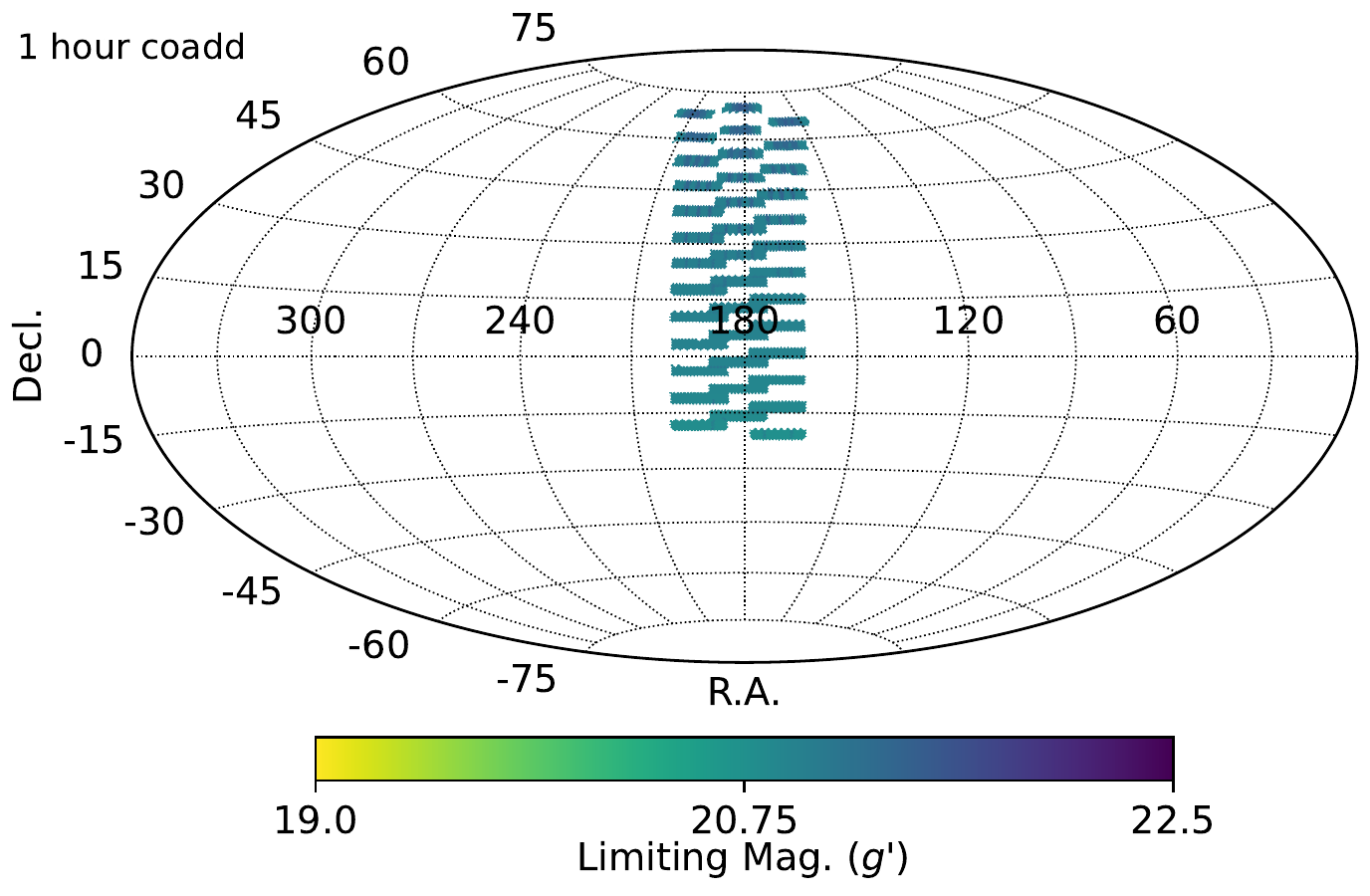}
     \end{subfigure}
     \hfill
     \begin{subfigure}
         \centering
         \includegraphics[width=0.48\textwidth]{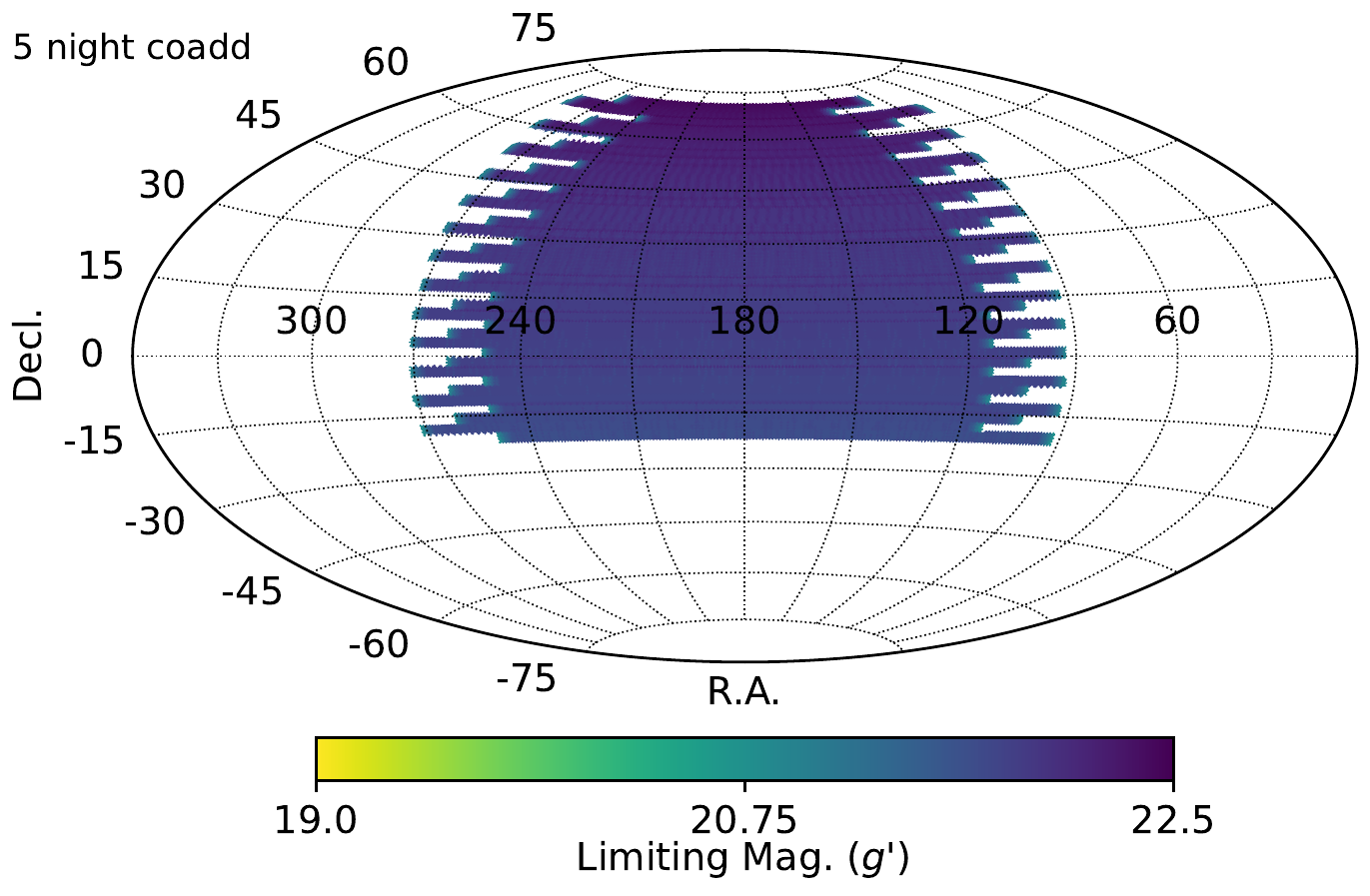}
     \end{subfigure}
    \caption{\textbf{Top left:} A rendering of the Argus Pathfinder mount. Pathfinder is a prototype system for the full Argus Array, containing 38 telescopes mounted in a shared hemispherical bowl. The telescopes are arranged in rows of constant R.A. with telescopes in each row spaced apart in dec.\ by three FoVs. The rows are evenly spatially separated, to allow the telescopes to fit in the bowl. This pattern reduces the size of the bowl required to view a stripe of the sky between decs.\ -20$^{\circ}$ and 72$^{\circ}$ by observing the stripe non-continuously. The bowl tracks the sky during 15 minute ratchets, before ratcheting back to start a new set of observations. Not shown is the weather-sealed container that Pathfinder will be placed in. 
    \textbf{Top right and bottom row:} The limiting magnitude and sky coverage of the Argus Pathfinder as observations build up over 60s, 1hr, and 5 nights of observations. Limiting magnitudes are for 5$\sigma$ signal-to-noise ratio, calculated for an 8-in-aperture telescope, and Monte-Carlo simulated for median conditions at a site similar to the Argus deployment site. For details on the instrumentation, such as sensor and telescope specifications, see Ref.~\citenum{law_2022}.}
    \label{fig:pathfinder}
\end{figure}

\textbf{Survey depth over time:} An important performance metric is depth of observations over time. The Evryscopes and Argus, for example, gain signal-to-noise by summing (``coadding'') the images from successive exposures over the course of many ratchets and observing nights. We can use the coverage evaluation algorithm to create grids representative of these exposures and ratchets. A summation of these grids shows the number of times a telescope views a specific part of the sky during that series of observations. By pairing this information with signal-to-noise calculations, we obtain the limiting magnitude of the array as a function of sky coverage. As an example, the survey depth as a function of coverage for the Argus Pathfinder is shown in Fig.~\ref{fig:pathfinder} as it builds coverage over a single exposure, an hour of observations, and five nights of coadds. 

\section{TELESCOPE PACKING}
\label{sec:packing}
There is no single optimal solution for packing rectangular telescope FoVs across the spherical sky. For the Argus Array, as with the Evryscopes, we reduce this parameter space by requiring the telescopes to form stripes of continuous declination. We arrange the stripes so that the resulting sky coverage is mostly free of gaps and contains slight overlaps, allowing for photometric solutions between the fields. The ``stripes pattern'' is used because it is a close match of the geometry of the rectangular telescope FoVs. It is also beneficial for data analysis because fields maintain roughly the same positioning in subsequent cameras from one ratchet to the next, at least in equatorial regions. Here, we describe the method for creating the stripes pattern, determine whether a hemispherical array shape is optimal for this pattern, and finally discuss how to split the array into several smaller subarrays with similar arrangements.

\subsection{Monolithic Mount}
\subsubsection{Packing Geometry}
\label{sec:geometry}
The stripes pattern used by the Evryscopes, and the planned Argus Array, is a simple overlapping grid of telescopes with slight adjustments in separations near the pole to remove gaps from distortion effects. Stripes begin at the greatest observable dec.\ from the survey deployment site, set either by the pole or the maximum allowable viewing airmass, and are separated in angular spacing by $\text{FoV}_y\times(1-F_y)$. Here, F$_y$ is the specified overlap in the dec.\ direction. The dec.\ of the last stripe is also set by the minimum allowable observing altitude. In each row, telescopes are separated by $\text{FoV}_x\times(1-F_x)\times\sec{\delta}$, where F$_x$ is the specified overlap in the R.A. direction and $\delta$ is the declination of the stripe. Around the pole, each row will contain only one or a few telescopes. 

The stripes extend in both the positive and negative R.A. direction starting from an R.A. of 180 degrees, which we have arbitrarily defined to be the local meridian and center of the array, down to the observing altitude cutoff. We center each stripe on the meridian (assuming the array is pointed directly overhead), such that the two central telescopes' overlap is centered at an R.A. of 180 degrees. This makes the mount symmetric, and allows for the array to be more easily divided between multiple mounts if necessary. Depending on the parameters of the array, this method may produce a grid with more telescopes than available in the array. Points closest to the horizon can then be chosen by hand to be eliminated from the grid.

Near the pole, the NCP for northern-hemisphere surveys and the south celestial pole (SCP) for southern surveys, packing becomes more difficult. The distortion associated with projecting the rectangular FoVs onto that part of the sky can create gaps in sky coverage. To handle this, telescopes at decs.\ greater than $\sim$80 degrees require some manual adjustment. Gaps can be filled by creating more overlap in the y-direction. For the Argus Array, we arranged the rows above 80 degrees dec.\ so that they are separated by only 60\%-85\% of the FoV$_y$. A similar tactic can be used to achieve the same results by reducing the separation of the telescopes in R.A. Decreasing the separation of the telescopes does sacrifice total coverage area, so it is important to balance having small gaps with losing sky coverage due to large overlaps.

\subsubsection{Optimizing the Array Shape}
\label{sec:shape}
Using the Nelder-Mead optimization method, we evaluated using a non-spherical array shape. We hypothesized that telescopes with rectangular FoVs arranged into a smaller ellipsoidal array shape might provide similar sky coverage and overlap as a larger spherical arrangement. For the optimization, we began with the stripes pattern of packing telescopes in a hemisphere as described above, but allowed the telescopes pointings to vary as the shape evolved from spherical to ellipsoidal. We optimized for sky coverage and overlaps and found no significant decrease in arrangement size for similar coverage metrics. While in some configurations it helped reduce coverage gaps near the pole, the ellipsoidal mounts caused distortions that often left gaps in the sky coverage in central-sky regions. We conclude the best mounting array shape is hemispherical.

\subsubsection{Application: The Argus Array}
\label{sec:app}
\begin{figure}
     \centering
    \includegraphics[width=0.98\textwidth]{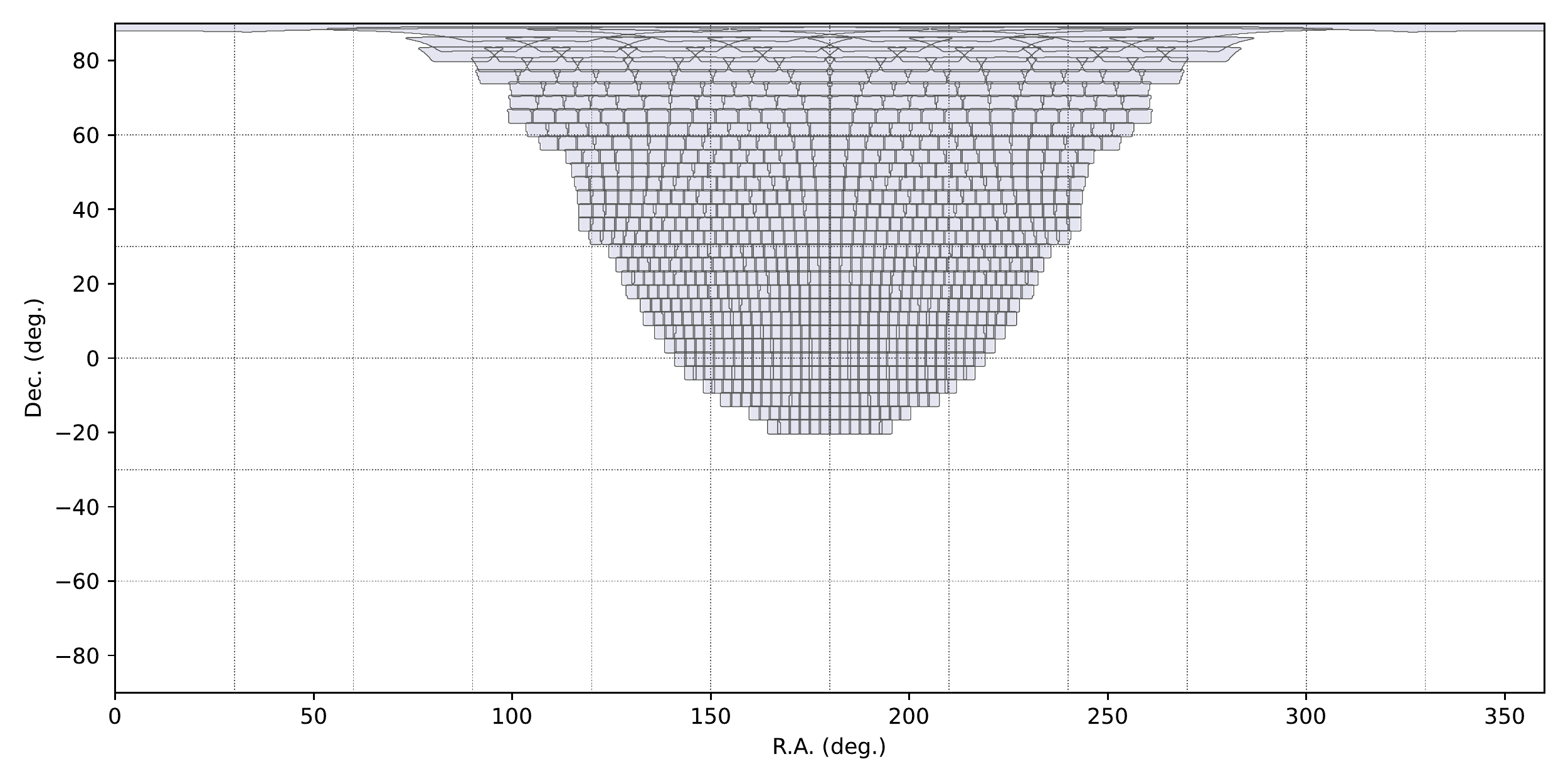}
    \caption{The Argus Array sky coverage pattern, designed for 900 8-in-aperture telescopes. The telescopes have a 3.4\% total overlap, 2.45 degree FoV in the x-direction, 3.67 degree FoV in the y-direction, and a total sky coverage of over 7,850 square degrees. Each shaded box represents an individual telescope's FoV on the sky. In the Argus hemispherical design\cite{law_2022}, this array would fit into a hemisphere of diameter $\sim$45 ft. In the new Argus pseudofocal design\cite{argus_spie}, the array will fit into a bowl of diameter $\sim$32 ft.}
    \label{fig:coverage_example}
\end{figure}

We now explore the packing of the Argus Array. Argus will contain 900 8-in telescopes. Each telescope has a 2.45 deg.\ by 3.67 deg.\ FoV. We require $\sim$1\% overlap in both the dec.\ and R.A. direction between telescopes, and will have a maximum observing airmass of two. Argus will be a northern-hemisphere survey located at a latitude above 30 degrees, and therefore will be able to observe the NCP. Using the stripes pattern outlined above, and manually adjusting pointings near the pole to minimize gaps, we achieve a total sky coverage of over 7,850 square degrees with 3.4\% total overlaps (Fig.~\ref{fig:coverage_example}). By placing telescopes in a hemispherical array, and requiring that they point normal to the spherical surface formed by their apertures, we find the minimum radius defining the arrangement to be $r\approx\frac{d}{\text{FoV}_\text{min}}=21.5$ ft, where \textit{d} is the size of the telescope tube diameter and $\text{FoV}_\text{min}$ is the smaller of the two FoV angles. A structural mount holding the telescopes in a hemisphere\cite{law_2022} must be at least a few feet larger than this radius, although the new Argus pseudofocal bowl design\cite{argus_spie} would be somewhat smaller because it does not require structure extending over the full hemisphere.

\subsection{Multiple Mount Configurations}
\label{sec:multi}
\begin{figure}
     \centering
    \includegraphics[width=0.98\textwidth]{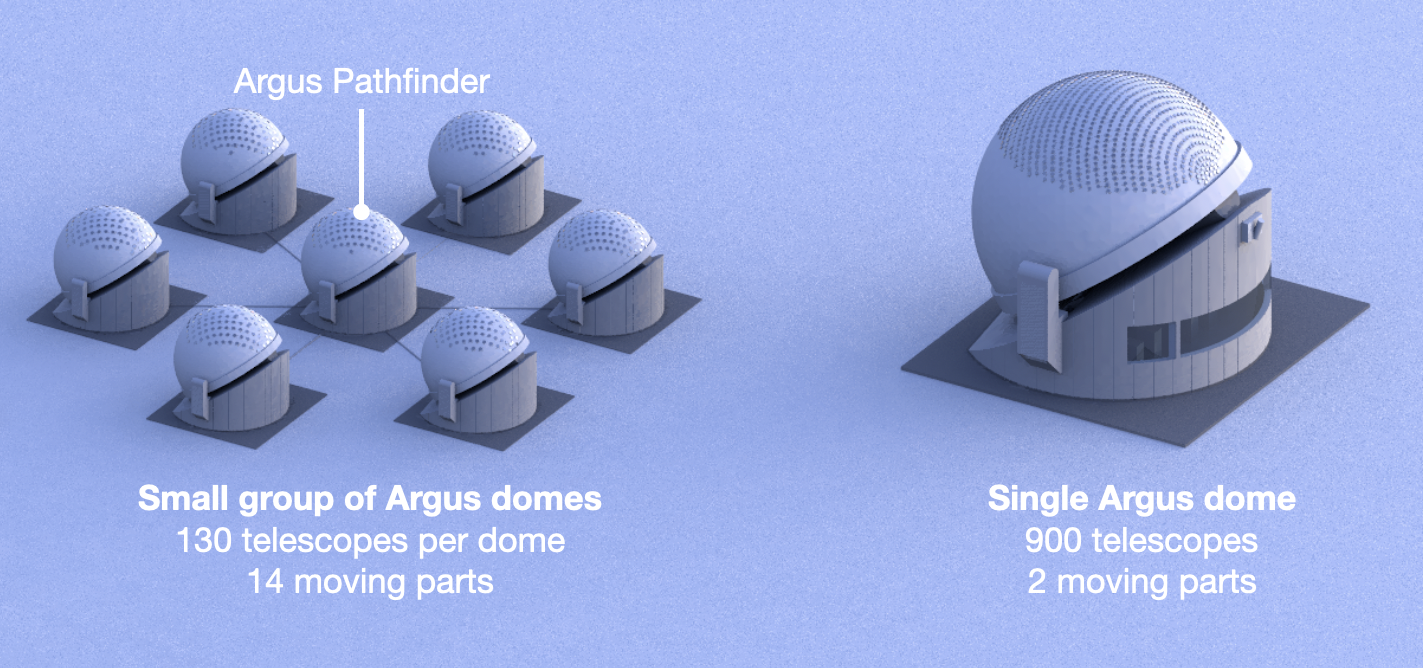}
    \caption{A visual comparison between a single mount and the multi-mount configuration. This shows the Argus Array concept placed into domes, instead of the pseudofocal design. The multi-mount configuration reduces the size of the required dome by half, at the cost of having an additional six mounts to maintain. This configuration has the benefit of allowing a prototype system, in this example Pathfinder, to serve as one of the subarrays when the full system is built. This figure is replicated from Ref.~\citenum{law_2022}.}
    \label{fig:arr_comp}
\end{figure}

When splitting the Argus Array into several smaller subarrays, we require that each subset contain similar numbers of telescopes and produce similar on-sky coverages. The method detailed below requires testing different numbers of mounts and determining the optimal value for one key parameter, the ``pattern constant''. This parameter, when specified for a set number of N subarrays, determines the pattern with which telescopes are placed into subarrays, and is described in more detail below. 

We start with the same sky packing as the monolithic array, and create a list of values ranging from 1 to N (1,2,...,N), representing the possible subarrays telescopes can be placed into. This list will be updated throughout the following algorithm. The first value in this list is the ``central subarray.'' This value sets which subarray the first telescope from each row will be placed into. Starting with the lowest declination stripe, the algorithm proceeds as follows:

\begin{enumerate}
    \item Place the first telescope on the positive side of the meridian into the central subarray for that row. 
    \item Move across the row in the direction of positive R.A., placing subsequent telescopes into subsequent subarrays until each telescope on the positive side of the meridian is placed. When all subarrays have one telescope from this row, start back at the beginning of the subarray list, placing the next telescope into subarray 1.
    \item Mirror this pattern across the meridian. Every telescope in this dec.\ stripe should now be in a subarray.
    \item Move up to the next row of dec.\ and iterate through the subarray list by a number of times given by the pattern constant, wrapping the list when necessary. This updated list now defines the new value for the central subarray. For example, with 5 mounts using a pattern constant of 2, the first telescope in the second dec.\ row would be placed in subarray 3. The following row would be started by placing the first telescope into subarray 5, and the row above that would start by placing a telescope into subarray 2.
    \item Repeat steps 1-4 for all declination stripes until every telescope is distributed into a subarray.
\end{enumerate}

\begin{figure}
     \centering
     \begin{subfigure}
         \centering
         \includegraphics[width=.65\linewidth]{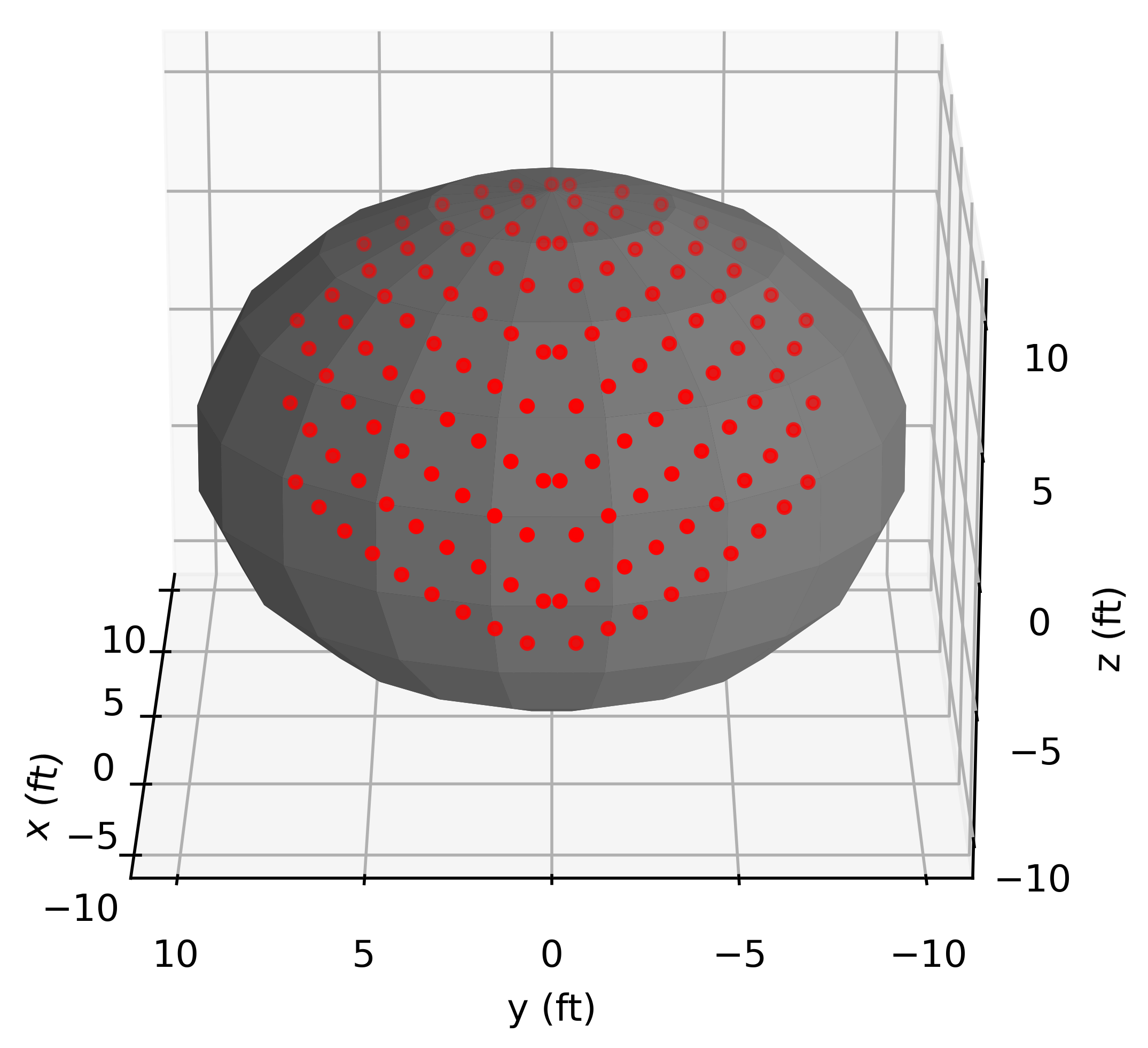}
     \end{subfigure}
     \begin{subfigure}
         \centering
         \includegraphics[width=.98\linewidth]{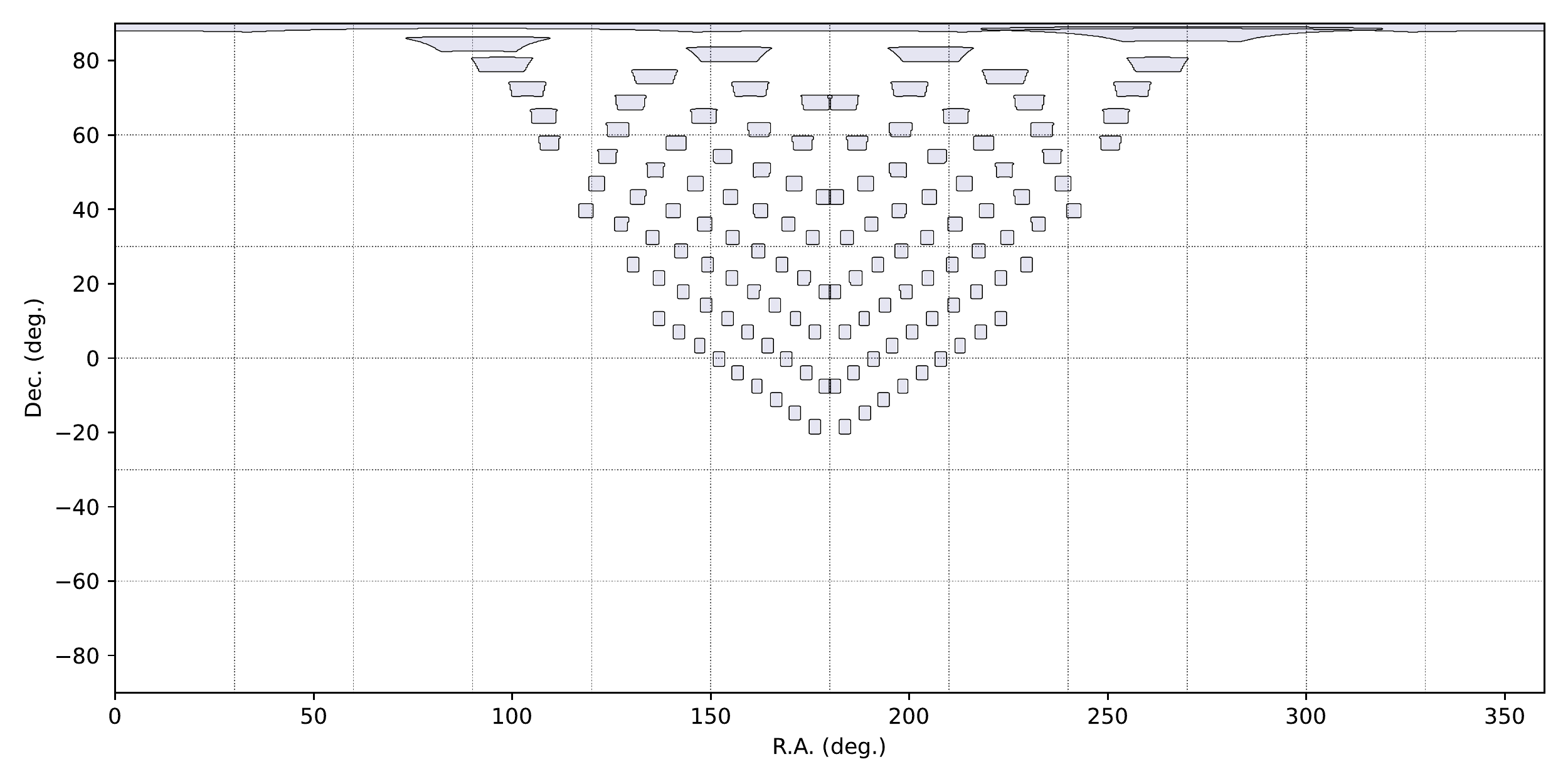}
     \end{subfigure}
     \hfill
    \caption{\textbf{Top:} 3D spatial distribution plot for one of the seven mounts in the Argus multi-mount configuration, oriented such that the z-axis is aligned with the celestial pole. Each red dot represents the location of one telescope in the arrangement. The telescopes in the middle, closest to the meridian, are too close to one another to fit in a standard hemispherical mount. This can be solved using either a pill shaped mount, or manually moving the telescopes to the open spaces above and below their current mounting position and manually re-pointing them to achieve the correct on-sky coverage.
    \textbf{Bottom:} The resulting sky coverage for the telescope placement pattern for one of the mounts in the Argus multi-mount configuration. It has non-continuous coverage for the entire dec.\ range of the system, allowing each subarray to observe the entire visible sky each night. }
    \label{fig:multi_dome}
\end{figure}

For the Argus Array, we found that using seven mounts with a pattern constant offset of five allowed us to use significantly smaller arrangements, with radii of 10 ft. Each dome would contain $\sim$130 telescopes. A comparison between the single- and multi-mount configurations can be seen in Fig.~\ref{fig:arr_comp}. For the Argus pattern, each mount also contained at least one telescope at every dec.\@, except near the pole where one telescope was placed manually, which allows each mount to get coverage of the entire sky over the observing night (Fig.~\ref{fig:multi_dome}).

\section{CONCLUSIONS}
\label{sec:conc}
Deployment of large optical arrays, such as the Argus Array, are becoming more feasible with recent advancements in commercially available astronomy equipment. These optical arrays will contain hundreds, or even thousands, of medium-aperture telescopes. Determining the best arrangement of these arrays is challenging, and requires large mounting structures. In this work, we explore the methods used to optimize telescope packing for the Argus Array in both single- and multiple-mount configurations. We develop an algorithm for rapidly evaluating the resulting sky coverage performance metrics. This algorithm is used to determine the total array FoV, overlap fraction, and create plots of sky coverage. With these metrics we can rapidly test different packing layouts and input parameters. We develop a method for packing telescopes in a hemispherical array, which we confirm to be the optimal arrangement. For the Argus Array we achieve continuous sky coverage of over 7,850 square degrees, with few-percent overlaps in coverage between telescopes, using a mount spanning $\sim$45 ft in diameter.

We find that by distributing the array over seven mounts we can reduce the size of each subarray by half, while allowing each subarray to observe the entire sky over the course of a single night. Building several smaller mounting structures could prove to be more cost-effective, at the cost of adding complexity. The reduced size of the structures could allow the multi-mount configuration to be constructed at sites with limited accessibility, where building a monolithic array may not be possible. Perhaps the most important benefit of a multi-mount array is that it enables a ``full-sized'' prototype system to be initially constructed, and completion of the array only requires replicating the prototype several times over. The Argus Pathfinder will help determine whether this strategy should be followed when building the full Argus Array.

\acknowledgments

This paper was supported by NSF MSIP (AST-2034381) and by the generosity of Eric and Wendy Schmidt by recommendation of the Schmidt Futures program. This research, and the construction of the Argus prototypes, is undertaken with the collaboration of the Be A Maker (BeAM) network of makerspaces at UNC Chapel Hill and the UNC BeAM Design Center.

\bibliography{report} 

\begin{thebibliography}{10}

\bibitem{law_2015}
{Law}, N.~M., {Fors}, O., {Ratzloff}, J., {Wulfken}, P., {Kavanaugh}, D.,
  {Sitar}, D.~J., {Pruett}, Z., {Birchard}, M.~N., {Barlow}, B.~N., {Cannon},
  K., {Cenko}, S.~B., {Dunlap}, B., {Kraus}, A., and {Maccarone}, T.~J.,
  ``{Evryscope Science: Exploring the Potential of All-Sky Gigapixel-Scale
  Telescopes},'' {\em \pasp}~{\bf 127},  234 (Mar. 2015).

\bibitem{ratzloff_2019}
{Ratzloff}, J.~K., {Law}, N.~M., {Fors}, O., {Corbett}, H.~T., {Howard}, W.~S.,
  {del Ser}, D., and {Haislip}, J., ``{Building the Evryscope: Hardware Design
  and Performance},'' {\em \pasp}~{\bf 131},  075001 (July 2019).

\bibitem{ratzloff2019variables}
{Ratzloff}, J.~K., Corbett, H.~T., Law, N.~M., Barlow, B.~N., Glazier, A.,
  Howard, W.~S., Fors, O., del Ser, D., and Trifonov, T., ``Variables in the
  southern polar region evryscope 2016 data set,'' {\em \pasp}~{\bf 131}(1002),
   084201 (2019).

\bibitem{ratzloff_2020_hot}
{Ratzloff}, J.~K., {Barlow}, B.~N., {N{\'e}meth}, P., {Corbett}, H.~T.,
  {Walser}, S., {Galliher}, N.~W., {Glazier}, A., {Howard}, W.~S., and {Law},
  N.~M., ``{Hot Subdwarf All Southern Sky Fast Transit Survey with the
  Evryscope},'' {\em \apj}~{\bf 890},  126 (Feb. 2020).

\bibitem{galliher_2020}
{Galliher}, N.~W., {Ratzloff}, J.~K., {Corbett}, H., {Law}, N.~M., {Howard},
  W.~S., {Glazier}, A.~L., {Vasquez Soto}, A., and {Gonzalez}, R.,
  ``{Evryscope-South Survey of Upper- and Pre-main Sequence Solar Neighborhood
  Stars},'' {\em \pasp}~{\bf 132},  114202 (Nov. 2020).

\bibitem{howard_superflare}
{Howard}, W.~S., {Tilley}, M.~A., {Corbett}, H., {Youngblood}, A., {Loyd},
  R.~O.~P., {Ratzloff}, J.~K., {Law}, N.~M., {Fors}, O., {del Ser}, D.,
  {Shkolnik}, E.~L., {Ziegler}, C., {Goeke}, E.~E., {Pietraallo}, A.~D., and
  {Haislip}, J., ``{The First Naked-eye Superflare Detected from Proxima
  Centauri},'' {\em \apjl}~{\bf 860},  L30 (June 2018).

\bibitem{howard_evryflare_1}
{Howard}, W.~S., {Corbett}, H., {Law}, N.~M., {Ratzloff}, J.~K., {Glazier}, A.,
  {Fors}, O., {del Ser}, D., and {Haislip}, J., ``{EvryFlare. I. Long-term
  Evryscope Monitoring of Flares from the Cool Stars across Half the Southern
  Sky},'' {\em \apj}~{\bf 881},  9 (Aug. 2019).

\bibitem{howard_evryflare_2}
{Howard}, W.~S., {Corbett}, H., {Law}, N.~M., {Ratzloff}, J.~K., {Galliher},
  N., {Glazier}, A., {Fors}, O., {del Ser}, D., and {Haislip}, J.,
  ``{EvryFlare. II. Rotation Periods of the Cool Flare Stars in TESS across
  Half the Southern Sky},'' {\em \apj}~{\bf 895},  140 (June 2020).

\bibitem{howard_evryflare_3}
{Howard}, W.~S., {Corbett}, H., {Law}, N.~M., {Ratzloff}, J.~K., {Galliher},
  N., {Glazier}, A.~L., {Gonzalez}, R., {Vasquez Soto}, A., {Fors}, O., {del
  Ser}, D., and {Haislip}, J., ``{EvryFlare. III. Temperature Evolution and
  Habitability Impacts of Dozens of Superflares Observed Simultaneously by
  Evryscope and TESS},'' {\em \apj}~{\bf 902},  115 (Oct. 2020).

\bibitem{glazier_trappist}
{Glazier}, A.~L., {Howard}, W.~S., {Corbett}, H., {Law}, N.~M., {Ratzloff},
  J.~K., {Fors}, O., and {del Ser}, D., ``{Evryscope and K2 Constraints on
  TRAPPIST-1 Superflare Occurrence and Planetary Habitability},'' {\em
  \apj}~{\bf 900},  27 (Sept. 2020).

\bibitem{corbett_2020}
{Corbett}, H., {Law}, N.~M., {Soto}, A.~V., {Howard}, W.~S., {Glazier}, A.,
  {Gonzalez}, R., {Ratzloff}, J.~K., {Galliher}, N., {Fors}, O., and {Quimby},
  R., ``{Orbital Foregrounds for Ultra-short Duration Transients},'' {\em
  \apjl}~{\bf 903},  L27 (Nov. 2020).

\bibitem{law_2022}
{Law}, N.~M., {Corbett}, H., {Galliher}, N.~W., {Gonzalez}, R., {Vasquez}, A.,
  {Walters}, G., {Machia}, L., {Ratzloff}, J., {Ackley}, K., {Bizon}, C.,
  {Clemens}, C., {Cox}, S., {Eikenberry}, S., {Howard}, W.~S., {Glazier}, A.,
  {Mann}, A.~W., {Quimby}, R., {Reichart}, D., and {Trilling}, D., ``{Low-cost
  Access to the Deep, High-cadence Sky: the Argus Optical Array},'' {\em
  \pasp}~{\bf 134},  035003 (Mar. 2022).

\bibitem{LAST}
{Ofek}, E.~O. and {Ben-Ami}, S., ``{Seeing-limited Imaging Sky
  Surveys{\textemdash}Small versus Large Telescopes},'' {\em \pasp}~{\bf 132},
  125004 (Dec. 2020).

\bibitem{argus_spie}
{Law}, N., {Corbett}, H., {Galliher}, N., {Gonzalez}, R., {Machia}, L.,
  {Vasquez Soto}, A., and {Walters}, G., ``{The Argus Optical Array: low-cost
  access to the deep, high-cadence sky},'' in [{\em Ground-based and Airborne
  Telescopes IX}{\nolinebreak\hspace{0.1em}]},  {\em \procspie} {\bf 12182}
  (July 2022).

\end{thebibliography}
\bibliographystyle{spiebib} 

\end{document}